\documentclass{svproc}
\usepackage{url}
\usepackage{amsmath}
\usepackage{graphicx}
\usepackage{array}
\usepackage{longtable}
\usepackage[table]{xcolor}
\usepackage{float}
\usepackage{comment}
\usepackage{bm}
\usepackage{cite}
\usepackage[colorlinks=true,    
            linkcolor=blue,     
            citecolor=blue,      
            urlcolor=black]   
           {hyperref}

\begin{document}
%
\title{Multimaterial topology optimization using SIMP, DMO and gSF: A comparative study with honeycomb tessellations}
\titlerunning{ MMTO using SIMP, DMO and gSF: A comparative study} 

\author{Duru Bhargav Kumar \and Prabhat Kumar}

\authorrunning{D.B. Kumar and P. Kumar}

\tocauthor{Duru Bhargav Kumar and Prabhat Kumar}


\institute{Indian Institute of Technology Hyderabad, Kandi, Telangana 502285, India \\ \email{pkumar@mae.iith.ac.in}}

\maketitle              

\begin{abstract}
This paper presents a comparative study of multimaterial topology optimization (MMTO) using the extended SIMP, Discrete Material Optimization (DMO), and generalized shape functions (gSF) approaches. The computational domain is discretized into hexagonal cells. Each cell is connected to its neighbors with edges. This characteristic reduces mesh-related effects on the optimized designs compared with conventional triangular and quadrilateral discretizations. Structural compliance is minimized at the prescribed volume fractions. The method of moving asymptotes is employed to update the design variables. The resulting optimized topologies are compared in terms of material distribution, structural performance, convergence behavior, and the characteristics of the obtained material interfaces. The comparative investigation provides insights into the working principles and performance of the extended SIMP, DMO, and gSF interpolation schemes for MMTO. 
\end{abstract}

\section{Introduction}
Topology optimization (TO) is a computational engineering technique used to identify an appropriate arrangement of material inside a prescribed design region under specified loading and support conditions. It seeks to optimize a selected objective function while satisfying the imposed geometric and physical restrictions~\cite{bendsoe2004topology}. This technique has been extensively applied to create lightweight and mechanically efficient structures by reducing compliance subject to resource limitations. In conventional single-material TO, the design process identifies the locations occupied by material and those left empty, assuming fixed material properties. In practical engineering applications, such as~\cite{Sigmund1997ThermalExpansion,kumar2022topologyMM,stegmann2005discrete,kumar2023topology,Gao2011MassConstraint,pinskier2024diversity,pinskier2023automated}, structural components may contain several material phases with distinct mechanical characteristics. This motivates the use of MMTO, which extends conventional TO to account for material selection with spatial arrangement~\cite{bendsoe2004topology,kumar2022topologyMM}. 

In MMTO, the optimization process determines both the location of material and the different candidate materials for each element (region) of the design domain. This makes the optimization problem more complex because different materials have varying properties, such as stiffness and density. To date, several interpolation schemes have been developed for MMTO~\cite{sarkar2026generalized}. Among them, the extended SIMP~\cite{Sigmund1997ThermalExpansion}, DMO~\cite{stegmann2005discrete,Gao2011MassConstraint}, and gSF~\cite{sarkar2026generalized} approaches provide different strategies for representing and distributing multiple materials. A comparative investigation of these approaches is therefore important for understanding their influence on the optimized topology, as performed in this paper.

Typically, the related boundary value problems are solved using the finite element (FE) method, which plays an important role in TO. Generally, triangular and quadrilateral elements are used~\cite{sarkar2025topology}. In this work, hexagonal finite elements are considered because each element has edge-connections with its neighboring elements. This feature naturally reduces mesh-related artifacts, including alternating element patterns and point-to-point connections, in the resulting optimized configurations~\cite{kumar2023honeytop90,kumar2015topology}.
Therefore, this study investigates MMTO with hexagonal finite elements. \texttt{HoneyMesher} provided with \texttt{HoenyTop90} MATLAB TO code~\cite{kumar2023honeytop90} is utilized to parametrize the design domain. Three material interpolation schemes, extended SIMP, DMO, and gSF, are implemented and evaluated comparatively. In the extended SIMP and DMO formulations, each element is characterized by $n$ design variables corresponding to the $n$ available materials, while the gSF approach provides $2^n-1$ possible material combinations. Compliance of the cantilever beam is minimized with the prescribed volume fractions for the candidate materials. The classical density filtering and projection scheme is employed. In the case of the gSF, the projection scheme is modified to cater to the design variable bounds~\cite{sarkar2026generalized}. We update the design variables using the method of moving asymptotes (MMA)~\cite{svanberg1987method}. The resulting topologies and material distributions are evaluated to investigate the differences in their optimization behavior and structural performance.

The paper is designed as follows. Section~\ref{sec2} describes the proposed methodology, covering the problem definition and optimization framework. Section~\ref{sec3} outlines the multimaterial interpolation strategies for extended SIMP, DMO, and gSF. Section~\ref{sec4:RD} presents the comparative results and their analysis. Finally, Section~\ref{sec5:con} summarizes the main findings and conclusions.

\section{Methodology}\label{sec2}
This section outlines the considered methodology, wherein the problem statement, TO formulation, and density and projection filtering are presented.

\subsection{Problem statement}

\begin{figure}[htbp]
	\centering
	\includegraphics[width=0.85\textwidth]{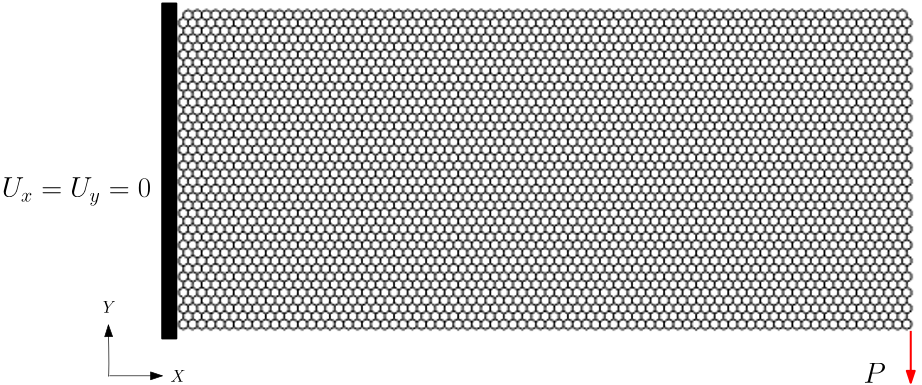}
	\caption{Cantilever-beam configuration discretized using a hexagonal finite-element mesh, with the left edge clamped and a point force applied at the bottom-right corner. The notation $U_x$ and $U_y$ denotes the horizontal and vertical displacement components, respectively~\cite{kumar2023honeytop90}.}
	\label{fig:cantilever}
\end{figure}

The numerical study considers a two-dimensional cantilever beam shown in Fig.~\ref{fig:cantilever}. The beam is clamped at its left edge, while a concentrated force acts at the lower-right corner of the unsupported end. The computational domain is discretized with hexagonal finite elements. Several candidate materials with distinct mechanical properties are available for assignment throughout the design region. The optimization task is to identify a suitable material distribution that minimizes structural compliance while satisfying the prescribed material-volume constraints. To investigate the influence of material interpolation on the optimized designs, three approaches, SIMP, DMO, and gSF, are implemented, and the resulting structural configurations are compared.

\subsection{Optimization Problem}

The objective of the present study is to obtain a stiff structure with a
prescribed amount of various materials while minimizing the compliance. The optimization problem is formulated as
\begin{equation}
	\begin{array}{ll}
		\displaystyle \min_{\mathbf{x}} &
		C = \mathbf{F}^\top\mathbf{U} \\[2mm]
		\text{subject to} &\\
		&\bm{\lambda}:\,\mathbf{K}(\mathbf{x})\mathbf{U} = \mathbf{F}, \\[1mm]
		&
		\bm{\mu}:\,\displaystyle \frac{V(\mathbf{x})_m}{V_{0m}} \leq V_{fm},\,{m = 1,\,2,\cdots,\,M}\\[1mm]
		&
		0 \leq x_{e,m} \leq 1.
	\end{array}
	\label{eq:complete_optimization}
\end{equation}
where $V_{0m}$ denotes the initial domain volume associated with 
material $m$, and $V_{fm}$ represents its prescribed volume fraction. 
The variable $x_{e,m}$ is the design parameter assigned to material 
$m$ in element $e$. 
\subsubsection{Density Filtering and Heaviside Projection}
By virtue of the geometrical construction of the hexagonal elements, the checkerboard patterns and point connections are automatically subdued. However, the optimized results still exhibit mesh dependence~\cite {kumar2023honeytop90}. Therefore, to circumvent mesh dependence and ensure a minimum length scale in the optimized geometry, we use the classical density-filtering technique here. The filtered design variable
for element $e$ is calculated as
\begin{equation}
	\tilde{x}_{e,m}
	=
	\frac{
		\displaystyle\sum_{j\in\mathcal{N}_e}
		W_{ej}x_{j,m}
	}{
		\displaystyle\sum_{j\in\mathcal{N}_e}
		W_{ej}
	},
	\label{eq:density_filter}
\end{equation}
where $\tilde{x}_{e,m}$ denotes the filtered value for material $m$ at 
element $e$, while $x_{j,m}$ represents the unfiltered design 
parameter for the same material at neighboring element $j$. The 
coefficient $W_{ej}$ specifies the weight assigned to the interaction 
between elements $e$ and $j$. The set $\mathcal{N}_e$ contains the 
neighboring elements located within the specified filter radius.  

The optimized results obtained with filtering typically contain a huge number of gray elements. Therefore, to achieve a solution close to 0-1 and a clear material boundary between materials, we use the Heaviside projection scheme. The projected physical variable is determined as
\begin{equation}
	x_{e,m}^{\mathrm{phys}}
	=
	\frac{
		\tanh(\beta\eta)
		+
		\tanh\left[
		\beta\left(\tilde{x}_{e,m}-\eta\right)
		\right]
	}{
		\tanh(\beta\eta)
		+
		\tanh\left[
		\beta(1-\eta)
		\right]
	},
	\label{eq:heaviside_projection}
\end{equation}
where $\beta$ controls the sharpness of the projection and $\eta$ is the
projection threshold. Increasing $\beta$ gradually transforms the
intermediate density field into a more discrete distribution. $\eta =0.5$ is used. The physical design variables
$x_{e,m}^{\mathrm{phys}}$ obtained after filtering and projection are
subsequently used in the SIMP, DMO, and gSF\footnote{In case of gSF, Eq.~\ref{eq:heaviside_projection} is tailored accordingly. Please see Sec.~\ref{sec3}} material interpolation schemes.

\section{Multi-Material Formulation}\label{sec3}

In the present study, an MMTO framework is
employed to determine the optimized material distribution simultaneously with
structural topology within the design domain. Since different candidate
materials possess different mechanical properties, an appropriate material
interpolation scheme is required to represent the effective material
properties of each FE. Herein, the extended SIMP~\cite{Sigmund1997ThermalExpansion}, DMO~\cite{stegmann2005discrete,Gao2011MassConstraint} and gSF~\cite{sarkar2026generalized} interpolation schemes
are investigated and compared. For each interpolation scheme, the effective constitutive matrix of an
element is determined from the corresponding design variables and material
properties. The resulting material distribution is then used to construct
the global stiffness matrix and evaluate the structural compliance. The
three approaches are implemented using the same hexagonal FE
discretization, loading conditions, boundary conditions, and optimization
constraints, allowing a direct comparison of their resulting topologies
and material distributions. 

```latex
\subsubsection{Extended SIMP}

In the SIMP formulation, the element stiffness is interpolated
using a penalized density variable. The exponent $p$ penalizes
intermediate densities, encouraging distinct solid and void
regions. For a single material, the effective Young's modulus is
defined as~\cite{Sigmund1997ThermalExpansion}
\begin{equation}
	E_e=(1-x_e^p)E_{\min}+x_e^pE_1,
	\label{eq:simp_single}
\end{equation}
where $x_e$ is the element design variable, $E_1$ is the solid
material's Young's modulus, and $E_{\min}$ is a small residual
stiffness used to avoid numerical singularities. For two candidate materials, two design variables control the
solid occupancy and material selection. The effective modulus is
\begin{equation}
	\begin{aligned}
		E_e = &
		(1-x_{e1}^{p})E_{\min}
		+
		x_{e1}^{p}
		\left[
		(1-x_{e2}^{p})E_1
		+
		x_{e2}^{p}E_2
		\right],\\
		=&(1-x_{e1}^{p})E_{\min}
		+
		x_{e1}^{p}(1-x_{e2}^{p})E_1
		+
		x_{e1}^{p}x_{e2}^{p}E_2.
	\end{aligned}
	\label{eq:simp_two_material}
\end{equation}
where $x_{e1}$ and $x_{e2}$ are the interpolation variables,
and $E_1$ and $E_2$ are the moduli of the candidate materials. For $M$ candidate materials, the formulation extends to $M$
design variables per element, giving
\begin{equation}
	\footnotesize
	E_e ={}
	(1-x_{e1}^{p})E_{\min}
	+
	\sum_{m=1}^{M-1}
	\left[
	\left(
	\prod_{i=1}^{m}x_{ei}^{p}
	\right)
	(1-x_{e,m+1}^{p})E_m
	\right]
	+
	\left(
	\prod_{i=1}^{M}x_{ei}^{p}
	\right)E_M .
	\label{eq:simp_general}
\end{equation}
Here, $x_{em}$ is the design variable at interpolation level $m$,
and $E_m$ is the Young's modulus of material $m$.

\subsubsection{Discrete Material Optimization}

The DMO approach is employed to obtain
a discrete material distribution within the TO domain~\cite{stegmann2005discrete,Gao2011MassConstraint}.
Unlike conventional single-material TO, DMO enables
the optimization process to assign materials to each element of a set of candidates. Like in the SIMP formulation, a penalization exponent is
introduced to suppress intermediate material states and promote a
discrete distribution of the candidate materials.

For a two-material design, the effective elastic modulus of element $e$
is obtained as

\begin{equation}
	E_e =
	E_{\min}
	+
	(1-E_{\min})
	\left[
	x_{e1}^{p}(1-x_{e2}^{p})E_1
	+
	x_{e2}^{p}(1-x_{e1}^{p})E_2
	\right],
	\label{eq:dmo_two_material}
\end{equation}
For $M$ candidate materials, the DMO
Interpolation can be generalized as follows:
\begin{equation}
	E_e =
	E_{\min}
	+
	(1-E_{\min})
	\sum_{m=1}^{M}
	\left[
	x_{em}^{p}
	\prod_{\substack{j=1\\j\neq m}}^{M}
	(1-x_{ej}^{p})
	\right]E_m,
	\label{eq:dmo_general}
\end{equation}
$x_{em}$ denotes the element
design parameter corresponding to candidate material $m$, whereas
$E_m$ represents the normalized Young's modulus assigned to that
material. The symbol $M$ specifies the number of candidate materials
considered in the optimization. The expression
$
x_{em}^{p}
\prod_{\substack{j=1\\j\neq m}}^{M}
(1-x_{ej}^{p})$
acts as a material-selection function for candidate $m$.

\subsubsection{Generalized Shape Function (gSF)}
The gSF approach offers an alternative interpolation
framework for MMTO~\cite{sarkar2026generalized}. It
generalizes conventional 1D, 2D, and 3D
shape functions to an $n$D multilinear formulation while
preserving their barycentric properties.
With $n$ design variables assigned to each element, the framework
can provide up to $2^n$ material phases. The
shape functions determine the density contribution of
each candidate material within an element.

In the one-dimensional case, the formulation depends on a single
natural coordinate, $X_1$, and describes two material states.
The corresponding density functions are given by:
\begin{equation}\label{eq:gsf_1d_density}
	x_1=\frac{1-X_1}{2},\qquad x_2=\frac{1+X_1}{2}.
\end{equation}
These density functions satisfy $x_1+x_2=1$ with $x_1\ge 0$ and $x_2\ge 0$. Likewise, for a two-dimensional formulation with two natural coordinates, $X_1$ and $X_2$, the density functions are given by:
\begin{equation}
	\scriptsize
	x_1 = \frac{(1-X_1)(1-X_2)}{4},\, x_3 = \frac{(1+X_1)(1+X_2)}{4},\,
	x_2 = \frac{(1+X_1)(1-X_2)}{4},\,x_4 = \frac{(1-X_1)(1+X_2)}{4}.
	\label{eq:gsf_2d_density}
\end{equation}
In general, for $n$ natural coordinates, the number of material phases is $2^n$, where in the density function of  the $m$-th material state is written as:
\begin{equation}x_m =\prod_{i=1}^{n}\frac{1+s_{mi}X_i}{2},\qquad m=1,\ldots,2^n \label{eq:gsf_general_density}
\end{equation}
where $s_{mi}\in\{-1,+1\}$ defines $n$D hypercube vertices associated with the $m$-th material state. The effective Young's modulus is determined as:
\begin{equation}
	E_j =E_{\min}+\sum_{m=1}^{2^n}x_{mj}^{p}\left(E_m-E_{\min}\right),\label{eq:gsf_general_interpolation}
\end{equation}
The effective modulus of element $j$, denoted by $E_j$, is
determined by the material property $E_m$ and the corresponding
element density variable $x_{mj}$. 

In the gSF framework, applying a
filter to the material density fields produces the same result
as filtering the underlying design variables. Therefore, the
standard density-filtering procedure remains applicable without
any changes. The projection operation is modified as~\cite{sarkar2026generalized}.

\begin{equation}
	x_{e,m}^{\mathrm{phys}} =
	\frac{
		2\tanh(\beta\eta)
		+
		\tanh\!\left(
		\beta\left(\frac{\tilde{x}_{e,m}+1}{2}-\eta\right)
		\right)
	}{
		\tanh(\beta\eta)
		+
		\tanh\!\left(\beta(1-\eta)\right)
	}
	-1
\end{equation}

The sensitivity of the compliance objective is derived through
the adjoint-variable method, with the chain rule used to account for
the objective's dependence on the design variables. Readers are referred to~\cite{sarkar2026generalized} for a complete derivation.

\begin{table}[H]
	\centering
	\caption{Material phases and corresponding colors.}
	\label{tab:material_colors}
	\begin{tabular}{c c@{\qquad\qquad}c c}
		\hline
		\textbf{Phase} & \textbf{Color} &
		\textbf{Phase} & \textbf{Color} \\
		\hline
		
		1 & \cellcolor{black}
		&
		9 & \cellcolor{cyan!50} \\
		
		2 & \cellcolor{cyan}
		&
		10 & \cellcolor{red!50!black} \\
		
		3 & \cellcolor{magenta}
		&
		11 & \cellcolor{orange} \\
		
		4 & \cellcolor{yellow}
		&
		12 & \cellcolor{yellow!70!brown} \\
		
		5 & \cellcolor{green}
		&
		13 & \cellcolor{violet} \\
		
		6 & \cellcolor{blue}
		&
		14 & \cellcolor{green!70!yellow} \\
		
		7 & \cellcolor{red}
		&
		15 & \cellcolor{olive} \\
		
		8 & \cellcolor{green!50}
		&
		\textbf{Void} & $10^{-9}$ \\
		
		\hline
	\end{tabular}
\end{table}

\section{Results and Discussion}\label{sec4:RD}
The cantilever beam is solved with different numbers of candidate materials using the extended  SIMP, DMO, and gSF
interpolation schemes, and their results are compared. All the optimization parameters are kept the same. \texttt{HoneyMesher} is used to describe the cantilever beam, wherein $\mathtt{HNelx} =120$ and $\mathtt{HNey}=60$ hexagonal FEs are used in the horizontal and vertical directions, respectively. The projection parameter starts with 1 and increases up to $\beta_{\max}=32$. The filter radius $r_{\mathrm{fill}}=4$ is set. The normalized Young's modulus assigned to the $i$-th material phase
is defined as
\begin{equation}
	\bar{E}_{0,i}
	=
	\frac{i-1}{N_{\mathrm{MP}}-1},
	\qquad
	i=1,2,\ldots,N_{\mathrm{MP}},
	\label{eq:E0}
\end{equation}
$N_{\mathrm{MP}}$ specifies the overall number of material phases
considered. The color assignments and corresponding
normalized Young's moduli are presented in Table~\ref{tab:material_colors}.

\begin{longtable}{c|@{\hspace{3mm}}c@{\hspace{3mm}}c@{\hspace{3mm}}c}
	
	\caption{Optimized topologies obtained 
		using extended SIMP, DMO, and gSF for different material counts.
		$M$ denotes the number of candidate materials, and $nDV$
		the number of design variables per element.}
	\label{tab:topology_comparison}\\
	
	\hline
	\textbf{M}
	&
	\textbf{SIMP}
	&
	\textbf{DMO}
	&
	\textbf{gSF}
	\\
	\hline
	\endfirsthead
	
	\hline
	\textbf{M}
	&
	\textbf{SIMP}
	&
	\textbf{DMO}
	&
	\textbf{gSF}
	\\
	\hline
	\endhead
	
	1
	&
	\begin{minipage}{0.23\textwidth}
		\centering
		\includegraphics[width=\linewidth]{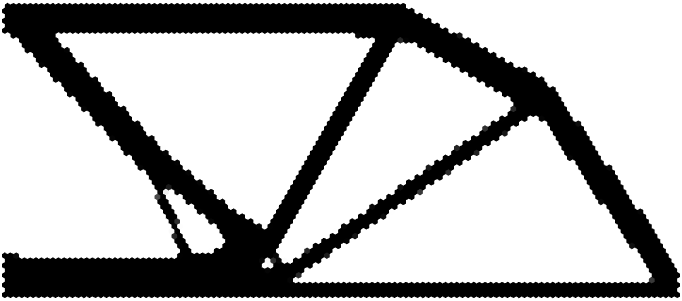}
		
		\vspace{1mm}
		$C = \mathrm{2.49342}$
		
		{nDV =1}
		
	\end{minipage}
	&
	\begin{minipage}{0.23\textwidth}
		\centering
		\includegraphics[width=\linewidth]{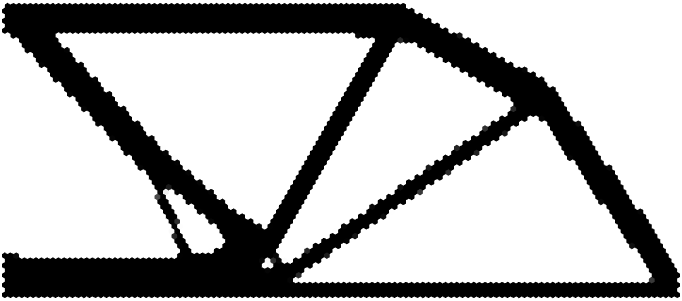}
		
		\vspace{1mm}
		$C = \mathrm{2.6527}$
		
		{nDV =1}
		
	\end{minipage}
	&
	\begin{minipage}{0.23\textwidth}
		\centering
		\includegraphics[width=\linewidth]{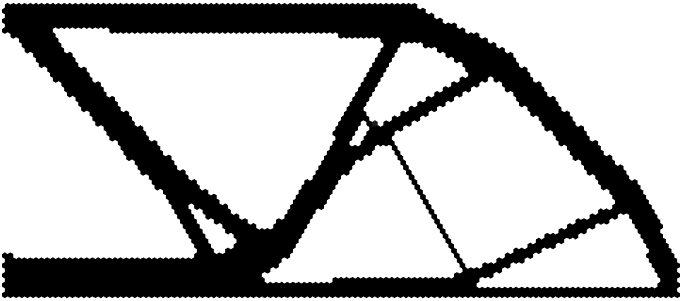}
		
		\vspace{1mm}
		$C = \mathrm{2.7522}$
		
		{nDV =1}
		
	\end{minipage}
	\\
	\hline
	
	2
	&
	\begin{minipage}{0.23\textwidth}
		\centering
		\includegraphics[width=\linewidth]{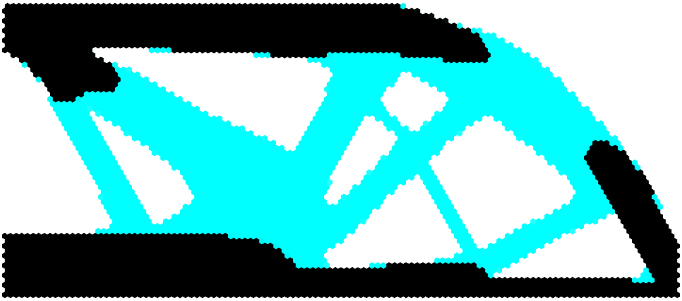}
		
		\vspace{1mm}
		$C = \mathrm{0.05061}$
		
		{nDV =2}
		
	\end{minipage}
	&
	\begin{minipage}{0.23\textwidth}
		\centering
		\includegraphics[width=\linewidth]{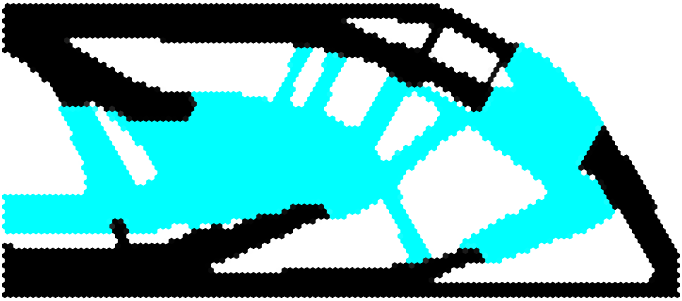}
		
		\vspace{1mm}
		$C = \mathrm{0.27068}$
		
		{nDV =2}
		
	\end{minipage}
	&
	\begin{minipage}{0.23\textwidth}
		\centering
		\includegraphics[width=\linewidth]{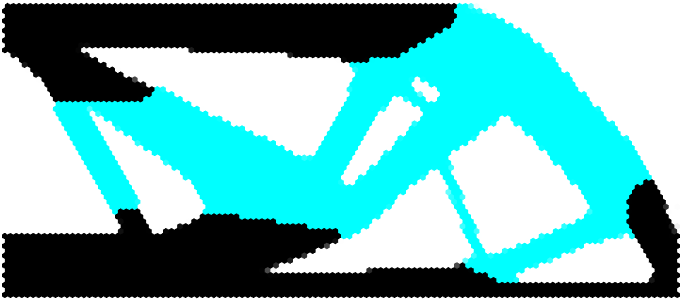}
		
		\vspace{1mm}
		$C = \mathrm{0.37425}$
		
		{nDV =2}
		
	\end{minipage}
	\\
	\hline
	
	3
	&
	\begin{minipage}{0.23\textwidth}
		\centering
		\includegraphics[width=\linewidth]{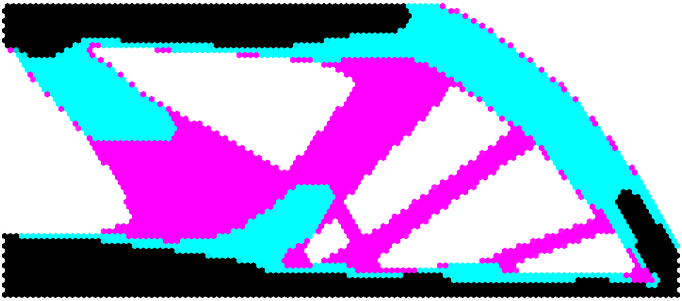}
		
		\vspace{1mm}
		$C = \mathrm{0.06532}$
		
		{nDV =3}
		
	\end{minipage}
	&
	\begin{minipage}{0.23\textwidth}
		\centering
		\includegraphics[width=\linewidth]{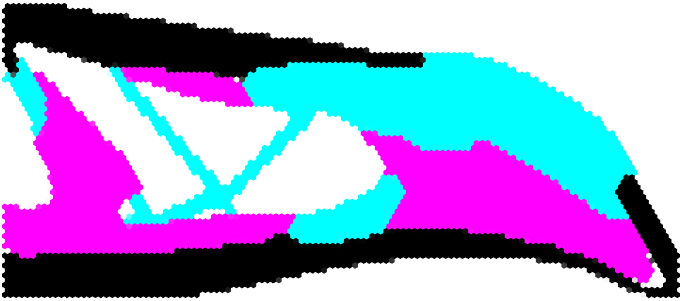}
		
		\vspace{1mm}
		$C = \mathrm{0.42766}$
		
		{nDV =3}
		
	\end{minipage}
	&
	\begin{minipage}{0.23\textwidth}
		\centering
		\includegraphics[width=\linewidth]{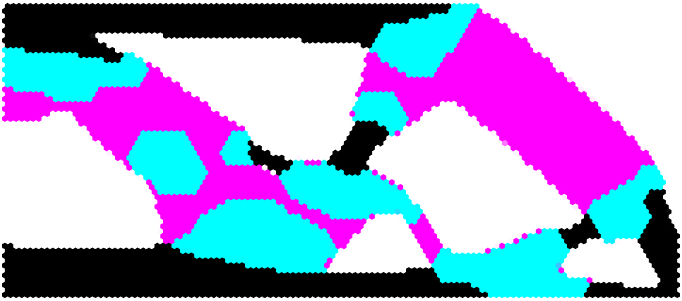}
		
		\vspace{1mm}
		$C = \mathrm{0.55503}$
		
		{nDV =2}
		
	\end{minipage}
	\\
	\hline
	
	4
	&
	\begin{minipage}{0.23\textwidth}
		\centering
		\includegraphics[width=\linewidth]{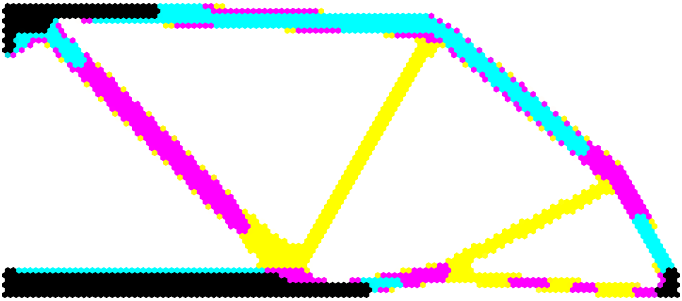}
		
		\vspace{1mm}
		$C = \mathrm{0.01492}$
		
		{nDV =4}
		
	\end{minipage}
	&
	\begin{minipage}{0.23\textwidth}
		\centering
		\includegraphics[width=\linewidth]{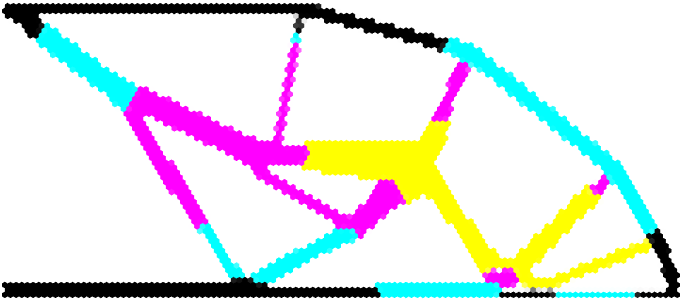}
		
		\vspace{1mm}
		$C = \mathrm{0.14289}$
		
		{nDV =4}
		
	\end{minipage}
	&
	\begin{minipage}{0.23\textwidth}
		\centering
		\includegraphics[width=\linewidth]{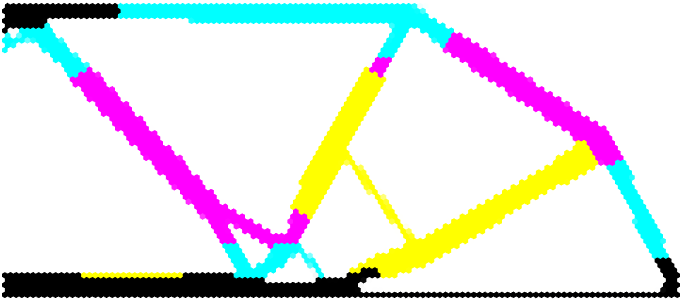}
		
		\vspace{1mm}
		$C = \mathrm{0.21971}$
		
		{nDV =2}
	\end{minipage}
	\\
	\hline
	
	5
	&
	\begin{minipage}{0.23\textwidth}
		\centering
		\includegraphics[width=\linewidth]{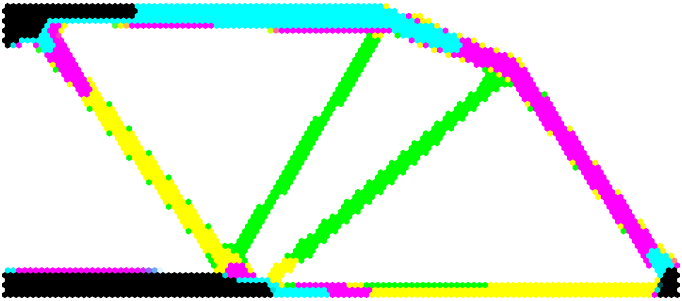}
		
		\vspace{1mm}
		$C = \mathrm{0.01280}$
		
		{nDV =5}
	\end{minipage}
	&
	\begin{minipage}{0.23\textwidth}
		\centering
		\includegraphics[width=\linewidth]{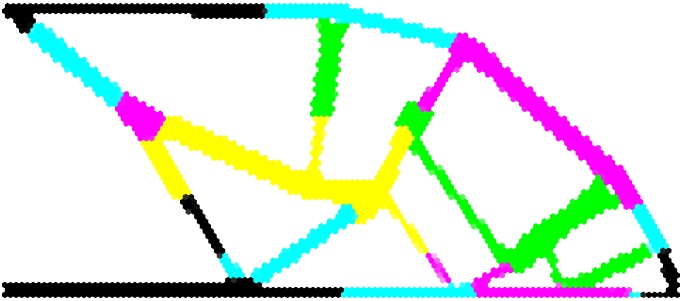}
		
		\vspace{1mm}
		$C = \mathrm{0.17688}$
		
		{nDV =5}
	\end{minipage}
	
	&
	\begin{minipage}{0.23\textwidth}
		\centering
		\includegraphics[width=\linewidth]{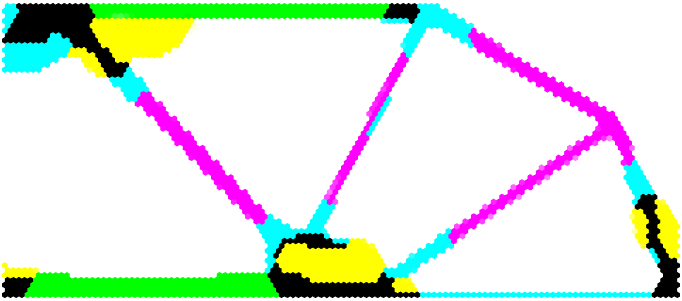}
		
		\vspace{1mm}
		$C = \mathrm{0.38735}$
		
		{nDV =3}
	\end{minipage}
	\\
	\hline
	
	6
	&
	\begin{minipage}{0.24\textwidth}
		\centering
		\textbf{---}
	\end{minipage}
	&
	\begin{minipage}{0.23\textwidth}
		\centering
		\includegraphics[width=\linewidth]{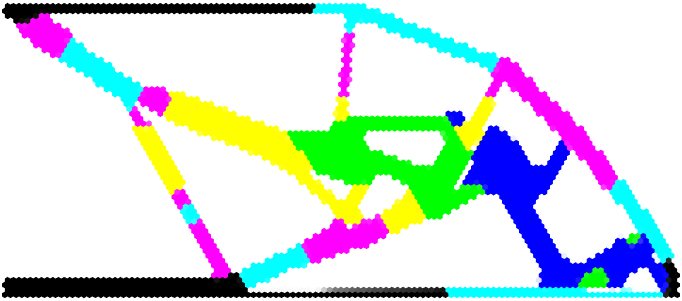}
		
		\vspace{1mm}
		$C = \mathrm{0.35165}$
		
		{nDV =6}
	\end{minipage}
	
	&
	\begin{minipage}{0.23\textwidth}
		\centering
		\includegraphics[width=\linewidth]{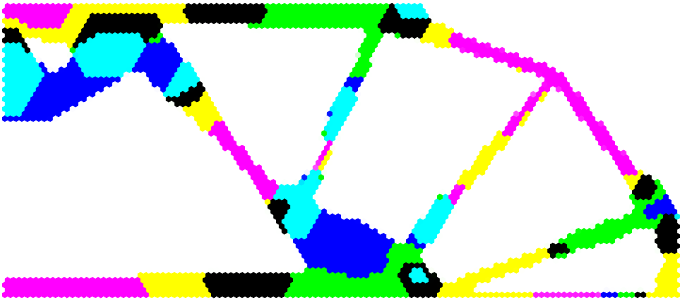}
		
		\vspace{1mm}
		$C = \mathrm{0.3517}$
		
		{nDV =3}
	\end{minipage}
	\\
	\hline
	
	7
	&
	\begin{minipage}{0.24\textwidth}
		\centering
		\textbf{---}
		
	\end{minipage}
	&
	\begin{minipage}{0.23\textwidth}
		\centering
		\includegraphics[width=\linewidth]{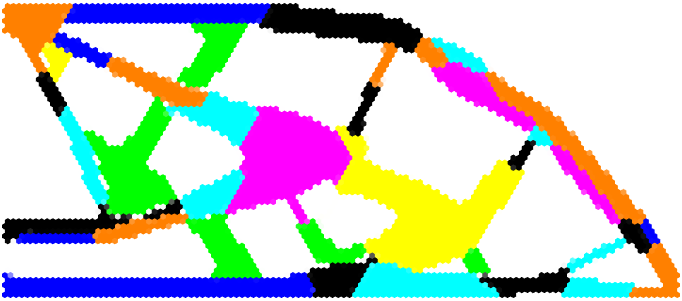}
		
		\vspace{1mm}
		$C = \mathrm{0.14161}$
		
		{nDV =7}
	\end{minipage}
	
	&
	\begin{minipage}{0.23\textwidth}
		\centering
		\includegraphics[width=\linewidth]{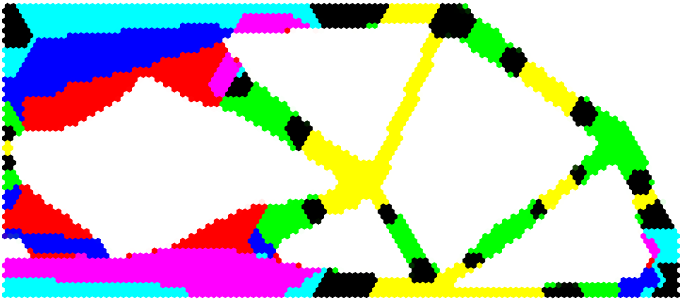}
		
		\vspace{1mm}
		$C = \mathrm{0.24724}$
		
		{nDV =3}
	\end{minipage}
	\\
	\hline
	
	8
	&
	\begin{minipage}{0.24\textwidth}
		\centering
		\textbf{---}
		
	\end{minipage}
	&
	\begin{minipage}{0.24\textwidth}
		\centering
		\textbf{---}
	\end{minipage}
	
	&
	\begin{minipage}{0.23\textwidth}
		\centering
		\includegraphics[width=\linewidth]{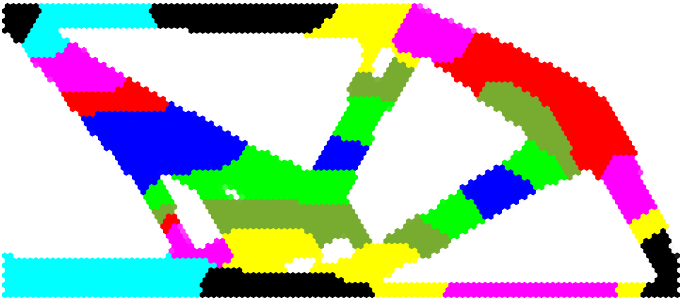}
		
		\vspace{1mm}
		$C = \mathrm{0.02922}$
		
		{nDV =3}
	\end{minipage}
	\\
	\hline
\end{longtable}

\begin{table}[htbp]
	\centering
	\caption{GSF results for higher numbers of materials.}
	\label{tab:GSF_high_materials}
	
	\begin{tabular}{c c c}
		\hline
		\textbf{M=9} & \textbf{M=10} & \textbf{M=11} \\
		\hline
		
		\begin{minipage}{0.28\textwidth}
			\centering
			\includegraphics[width=\linewidth]{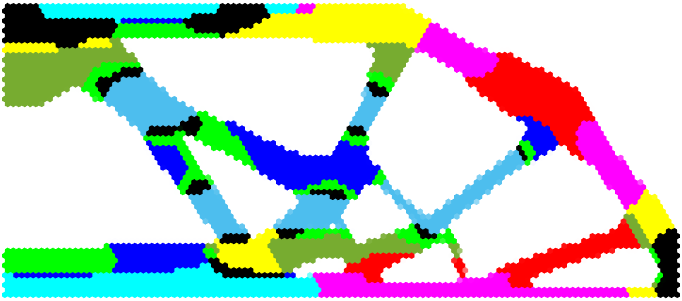}
			
			\vspace{1mm}
			$C = \mathrm{0.034534}$

		\end{minipage}
		&
		\begin{minipage}{0.28\textwidth}
			\centering
			\includegraphics[width=\linewidth]{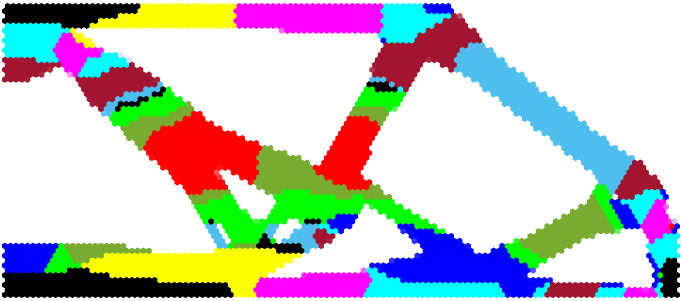}
			
			\vspace{1mm}
			$C = \mathrm{0.03963}$

		\end{minipage}
		&
		\begin{minipage}{0.28\textwidth}
			\centering
			\includegraphics[width=\linewidth]{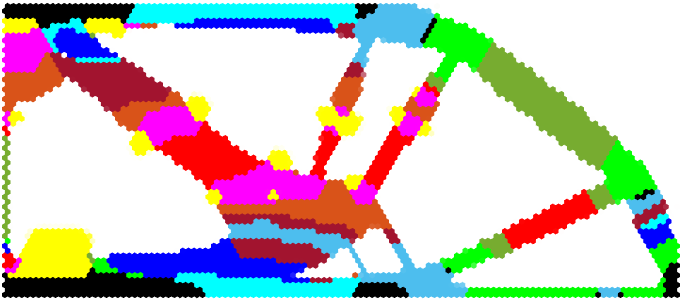}
			
			\vspace{1mm}
			$C = \mathrm{0.04424}$

		\end{minipage}
		\\[4mm]
		
		\hline
		
		\textbf{M=12} & \textbf{M=13} & \textbf{M=14} \\
		\hline
		
		\begin{minipage}{0.28\textwidth}
			\centering
			\includegraphics[width=\linewidth]{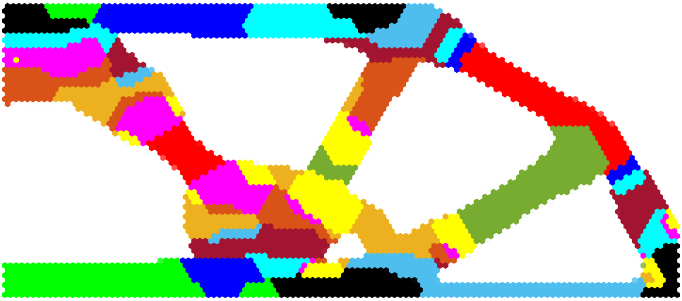}
			
			\vspace{1mm}
			$C = \mathrm{0.046770}$
			
		\end{minipage}
		&
		\begin{minipage}{0.28\textwidth}
			\centering
			\includegraphics[width=\linewidth]{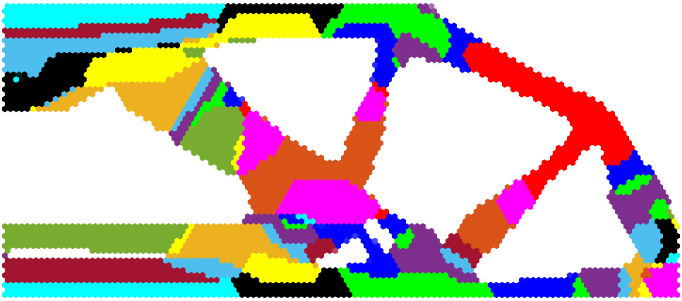}
			
			\vspace{1mm}
			$C = \mathrm{0.048369}$
			
		\end{minipage}
		&
		\begin{minipage}{0.28\textwidth}
			\centering
			\includegraphics[width=\linewidth]{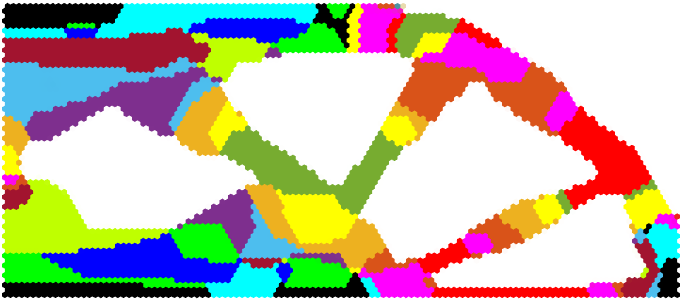}
			
			\vspace{1mm}
			$C = \mathrm{0.04272}$
			
		\end{minipage}
		\\[4mm]
		
		\hline
	\end{tabular}
\end{table}

Results with up to eight candidate materials are shown in Table.~\ref{tab:topology_comparison}. We note that the extended SIMP method fails to obtain optimized solutions for more than five candidate materials, wherein all volume constraints are active (or nearly active). Similarly, the DMO method fails beyond seven candidate materials for the design parameters considered. In contrast, the gSF approach can obtain optimized solutions for eight or more candidate materials with only a few design variables per element. The solutions for up to 14 materials (nDV =4) using the gSF approach are depicted in Table~\ref{tab:GSF_high_materials}. We also note that the optimized solutions from the gSF approach are competitive with those from the extended SIMP and DMO approaches.  
\section{Conclusion}\label{sec5:con}

The present study investigates MMTO using hexagonal finite elements.
Three interpolation schemes, the extended  SIMP, DMO, and
gSF, are implemented and compared for
different numbers of materials. The conclusions are summarized as
follows:

\begin{enumerate}
	\item The SIMP formulation effectively handled multi-material volume-fraction constraints for up to five materials with the employed set of parameters. However, an increase in the number of material phases leads to a corresponding rise in the number of design variables per element, thereby increasing computational cost.
	
	\item The DMO formulation successfully handled volume-fraction constraints for up to seven materials with the employed set of parameters. Compared to SIMP, DMO offers greater flexibility in discrete material selection; nevertheless, like the SIMP approach, the increasing number of material-related design variables can heighten computational demand.
	
	\item The gSF formulation successfully handled volume-fraction constraints for an arbitrary number of candidate materials, as demonstrated here for up to 15 material phases, with the employed set of parameters. This highlights the potential of the gSF approach for TO problems involving large material sets.
	
	\item The comparison indicates that interpolation schemes' capacity to represent multiple materials varies with the number of materials. While SIMP and DMO may be used for a relatively small number of materials, gSF provides greater versatility for both small- and large-scale multi-material systems with fewer design variables. 
	\item The use of hexagonal finite elements provides a robust alternative discretization scheme for MMTO problem settings.
\end{enumerate}


\section*{Acknowledgment}
PK thanks the Anusandhan National Research Foundation, India, for support under the MATRICS project, file number MTR/2023/000524. 


\end{document}